\documentclass[conference]{IEEEtran}

\usepackage{amsfonts,amsmath,amssymb,amsthm}
\usepackage{balance}
\usepackage{graphicx}
\usepackage{amsmath}
\usepackage{graphicx}
\usepackage{calc}
\usepackage{epstopdf}
\usepackage{subfigure}
\usepackage[T1]{fontenc}
\usepackage{amsfonts}
\usepackage{bm}
\usepackage{bbm}
 \usepackage[ruled,norelsize]{algorithm2e}
\usepackage{algorithmic}
\usepackage{verbatim}
\usepackage{amssymb}
\usepackage{amsthm}

\usepackage{color}
\usepackage{subfigure}
\usepackage{array}
\usepackage{cite}
\usepackage{stfloats}

\ifCLASSINFOpdf

\else

\fi

\begin{document}

\title{Resource Allocation for a Full-Duplex Base Station Aided OFDMA System}

\author{
\IEEEauthorblockN{Yang You\IEEEauthorrefmark{1},  Chong Qin\IEEEauthorrefmark{2} and Yi Gong\IEEEauthorrefmark{2}}

\IEEEauthorblockA{\IEEEauthorrefmark{1}School of Electrical Engineering, KTH Royal Institute of Technology, Stockholm, Sweden}

\IEEEauthorblockA{\IEEEauthorrefmark{2}Department of Electrical and Electronic Engineering, Southern University of Science and Technology, Shenzhen, China}

}



%
%
%
%
\maketitle
\begin{abstract}

Exploiting full-duplex (FD) technology on base stations (BSs) is a promising solution to enhancing the system performance. Motivated by this, we revisit a full-duplex base station (FD-BS) aided OFDMA system, which consists of one BS, several uplink/downlink users and multiple subcarriers. A joint 3-dimensional (3D) mapping scheme among subcarriers, down-link users (DUEs), uplink users (UUEs) is considered as well as an associated power allocation optimization. In detail, we first decompose the complex 3D mapping problem into three 2-dimensional sub ones and solve them by using the iterative Hungarian method, respectively. Then based on the Lagrange dual method, we sequentially solve the power allocation and 3-dimensional mapping problem by fixing a dual point. Finally, the optimal solution can be obtained by utilizing the sub-gradient method. Unlike existing work that only solves either 3D mapping or power allocation problem but with a high computation complexity, we tackle both of them and have successfully reduced computation complexity from exponential to polynomial order. Numerical simulations are conducted to verify the proposed scheme.

\end{abstract}

\section{Introduction}
 A communication system throughput can be highly improved by exploiting the FD technology. However, due to the self-interference (SI) limitations, most of the previous research were restricted to half-duplex (HD) systems. Recently, as mentioned in \cite{1}, FD communication came to stage since great progress had been made in developing SI cancellation schemes. In particular, FD relays are widely considered in cooperative communication systems, such as \cite{92} and \cite{93}. However, FD technology is not only applicable for the relay node, it can also be applied to the BS. Currently, there exists some little research work on FD-BS. For instance, \cite{95} investigates the extra benefits on freedom degrees of cellular systems brought by the FD-BS. \cite{96} considered the security problem of a FD-BS aided system, which is an attractive topic.

In order to optimize the system performance, a low-complexity resource allocation strategy needs to be investigated in the FD-BS scenario. There exists some work related with this but failing achieving the goal in some degree, e.g. \cite{97} and \cite{98}.  In \cite{97}, it considered a cellular system consisting of one FD-BS, several users and sub-channels. A joint user scheduling, power control and channel assignment scheme was taken into account, but no closed-form solution was provided. in \cite{98}, a potential solution for power allocation problem was given, but only considering the mapping problem between sub-channel and either DUE or UUE.

In this paper, we revisit the system model investigated in \cite{93} and \cite{98}, majoring in solving a low complexity 3-dimensional pairing problem among UUE, DUE and sub-channels. Besides, the power allocation problem is also taken into account. The contributions of this paper are twofold: (1) Solving the power allocation problem of a FD-BS aided OFDMA system. (2) Significant computation complexity reduction on the 3D pairing problem.

The rest of the paper is organized as follows: Section II presents the system model. Sections III proposes a low complexity method solving the binary assignment problem. Section IV provides the joint power allocation and binary assignment scheme. The numerical results are presented in Sections V. Finally, section VI draws the conclusion.

\section{System Model and Problem Formulation}

\subsection{System Model}
By revisiting the model of Fig.1 in \cite{97}, we consider a multi-user FD-BS OFDMA system and it consists of one FD-BS, M UUEs and N DUEs. The users are all in HD mode and equipped with single antenna. The total bandwidth is divided into $K$ rayleigh flat fading subchannels, and each of them $ k $ can be shared by only one DUE $ n $ and UUE $ m $. This 3-dimensional combination is represented as $(m,n,k)$. Due to the imperfect SI cancellation, the BS will suffer from the signal transmitted from the uplink user. And the remaining SI is generally modeled as an additive white Gaussian noise (AWGN).


\subsection{Problem Formulation}
Define $m\in\{1,....M\}$, $n\in\{1,....N\}$ and $k\in\{1,....K\}$ as the index of UUEs, DUEs, and Sub-channels, respectively. $P_{b,n}^k$ denotes the transmitted  power from BS to a DUE over subcarrier $ k $ and $P_{m,b}^k$ represents the transmitted power from UUE to BS over the same channel. $h_{m,b}^k$ and $h_{b,n}^k$ are the corresponding down-link and uplink channel coefficients. The received signals of BS and DUE can be expressed as:
\begin{equation} \label{eq:Eq1}
y_{m,b}^k=\sqrt{P_{m,b}^k} h_{m,b}^k x_m+Z_D+Z_B
\end{equation}
\begin{equation} \label{eq:Eq1}
y_{b,n}^k=\sqrt{P_{b,n}^k} h_{b,n}^k x_n+\sqrt{P_{m,b}^k} h_{m,n}^k x_m+Z_B
\end{equation}
Where $ x_m $ and $ x_n $ denote the transmitted symbol of the UUE and BS, respectively. Their average power are normalized as $ \bf{\sl{E}}\{\left| x_m \right|^2\}=\bf{\sl{E}}\{\left|x_m\right|^2\}=1$, where $ \bf{\sl{E}} $ denotes the mathematical expectation. $ Z_D \sim CN(0,\sigma_D^2) $ and $ Z_B \sim CN(0,\sigma_B^2) $ denote the self-interference and receiver noise, respectively. Then, we obtain the throughput of  down-link and uplink over sub-channel $ k $ :

\begin{equation} \label{eq:Eq1}
R_{m,b,k}^{up}=\log \left(1+\frac{P_{m,b}^k \left| h_{m,b}^k \right|^2}{\sigma_D^2+\sigma_B^2} \right)
\end{equation}

\begin{equation} \label{eq:Eq1}
R_{b,n,k}^{down}=\log \left(1+\frac{P_{b,n}^k \left| h_{b,n}^k \right|^2}{P_{m,b}^k \left| h_{m,n}^k \right|^2+\sigma_N^2} \right)%
\end{equation}
The sum rate of transmission pair $(m,n,k)$ is $ R_{m,n}^k $:
\begin{equation} \label{eq:Eq1}
R_{m,n}^k = R_{m,b,k}^{up} + R_{b,n,k}^{down}
\end{equation}
To maximize the system throughput, the joint 3D mapping and power allocation problem is now formulated as:
\begin{equation}
\begin{aligned}
&(P1):\mathop {\max }\limits_{\{p_{b,n}^k, p_{m,b}^k, X\}} \sum\limits_{m = 1}^M {\sum\limits_{n = 1}^N {\sum\limits_{k = 1}^K {x_{m,n}^kR_{m,n}^k} } }\\
&\qquad  \qquad ~~ s. t. ~~~c1:\sum_{n=1}^{N}\sum_{k=1}^K p_{b,n}^k \leq P_b \\
&\qquad  \qquad ~~~~~~~~~ c2:\sum_{k=1}^K P_{m,b}^k \leq P_m, \quad\forall m \\
&\qquad  \qquad ~~~~~~~~~  c3: X_{m,n}^k=\{0,1\} \\
&\qquad  \qquad ~~~~~~~~~  c4:\sum\limits_{m = 1}^M \sum\limits_{n = 1}^N x_{m,n}^k = 1,\quad\forall k\\
&\qquad  \qquad ~~~~~~~~~~~~~  \sum\limits_{m = 1}^M\sum\limits_{k = 1}^K {x_{m,n}^k} = 1,\quad\forall n \\
&\qquad  \qquad ~~~~~~~~~~~~~  \sum\limits_{n = 1}^N\sum\limits_{k = 1}^K {x_{m,n}^k} = 1, \quad\forall m
\end{aligned}
\end{equation}
Where $X$ is the $M\times N \times K$ 3-dimensional assignment matrix with $x_{m,n}^k=1$ if subchannel $k$ is assigned to UUE-DUE pair $(m,n)$, and $x_{m,n}^k=0$, otherwise. The constraint c4 follows the fact each subchannel can only be assigned to one UUE-DUE pair, and constraint c1, c2 indicate the individual power constraint for UUE and BS, respectively.

\section{low-complexity 3-dimensional binary assignment scheme}
By fixing the power allocation scheme, (P1) can be simplified as:
\begin{equation} \label{eq:Eq1}
\begin{aligned}
&\mathop{\max }\limits_{X}\sum\limits_{m = 1}^M \sum\limits_{n = 1}^N {\sum\limits_{k = 1}^K {x_{m,n}^k{R_{m,n}^k}^*} }\\
&~~ s.t.~~c3,c4
\end{aligned}
\end{equation}

where ${R_{m,n}^k}^*$ represents a fixed throughput of each $(m,n,k)$ pair. (7) is a NP-complete problem and the complexity of exhaustive search method related with it is in exponential order shown in \cite{10}. However, since it's proved in \cite{99}, the 3-dimensional mapping problem can be solved by using 2-dimensional Hungarian method in 5 iterations with near-optimal performance. And the complexity turns to be $O(5L^{3})$ where $ L=\max( M,N,K) $, which is polynomial and much lower than the optimal scheme.

In this case, we can decompose our 3D mapping problem into following three different 2D mapping subproblems and solve them iteratively,

1) 2D mapping between the UUE-DUE pair $(m,n)$ and Subchannel $k$, the solution is $X_{1}^*$.

2) 2D mapping between the UUE-Subchannel pair $(m,k)$ and DUE $n$, the solution is $X_{2}^*$.

3) 2D mapping between the DUE-subchannel pair $(n,k)$ and UUE $m$, the solution is $X_{3}^*$.

To initialize the scheme, set an arbitrary mapping matrix $X_{M,N,K}$ which satisfies constraint c3 and c4. Then, define a 2-dimensional index including indices of UUE $m$ and DUE $n$ be $\chi_{0}= \{(m,n)\mid x_{m,n}^k =1,\forall m,n,k\}$. Here, for simplicity, we use $d$ to represent the pair $(m,n)$, when $M\leq N$, $d=m$; and $d=n$ otherwise.

With the pair $(m,n)$, the two different dimensions UUE $m$ and DUE $n$ could be regarded as a joint dimension with index $d$, then we could adopt a 2-dimensional mapping matrix $T_{\min(M,N)\times K}= [{t_{d}}^k]$ with ${t_{d}}^k=1$ if subchannel $k$ is allocated to UUE-DUE pair $(m,n)$, otherwise ${t_{d}}^k=0$.

In this case, the mapping relationship shown in the initial mapping matrix $X_{M,N,K}$ could be demonstrated by the set $(\chi_{0},T_{0})$. And the problem turns to be searching an optimal subchannel assignment matrix ${T_{0}}^*$ for the current UUE-DUE pair set $\chi_{0}$, which can be formulated by the following,

\begin{equation} \label{eq:Eq1}
\begin{aligned}
&\mathop {\max }\limits_{{T_{0}}} \sum\limits_{d = 1}^{\min(M,N)} {\sum\limits_{k = 1}^K {T_{d}^k{R_{d,k}^*}} }\\
&~~~s.t. \sum\limits_{d = 1}^{\min(M,N)}{t_{d}^k = 1},\forall k;\\
&~\quad \qquad R_{d,k}^*= {R_{m,n}^k}^*, \forall (m,n)\in \chi_{0}
\end{aligned}
\end{equation}

The classic Hungarian method can be used to solve the above problem, yielding a new 3D mapping matrix $X_{1}^*$ which has the same mapping relationship with set $(\chi_{0},{T_{0}}^*)$. And this $X_{1}^*$ can be set as the initial mapping matrix for subproblem 2). By applying same strategy iteratively on the three subproblems, we can derive the optimal solutions for each iteration in the sequence $X_{1}^*\rightarrow X_{2}^*\rightarrow X_{3}^*\rightarrow X_{4}^*\rightarrow X_{5}^*$. $X_{5}^*$ can be chosen as the global optimal 3D binary mapping matrix.

\section{Joint power allocation and 3D mapping}

In this section, we further consider the power allocation problem. As the joint power allocation and binary assignment
problem is a non-convex mixed combinatorial problem, which is extremely complicated to solve directly. We propose the Lagrange dual method, which has been proved in \cite{11}, that for multi-carrier systems, the duality gap of a non-convex resource allocation problem is negligible when the number of subcarriers becomes sufficiently large.

To combine the integer constraint with power constraint, we define the following virtual power:

\begin{equation}
P_{m,n}^{k,up}=P_{m,b}^k x_{m,n}^k
\end{equation}
\begin{equation}
P_{m,n}^{k,down}=P_{b,n}^k x_{m,n}^k
\end{equation}
With (9) and (10), constraint c1 and c2 can be transformed to:
\begin{equation}
c5: \sum_{k=1}^K\sum_{n=1}^N P_{m,n}^{k,up}\leq P_m \qquad\forall m
\end{equation}
\begin{equation}
c6: \sum_{k=1}^K\sum_{n=1}^N\sum_{m=1}^MP_{m,n}^{k,down}\leq P_b
\end{equation}
(P1) can be transformed to:

\begin{equation} \label{eq:Eq1}
\begin{aligned}
&(P3): \mathop{\max }\limits_{{X,\overline{P}}}\sum\limits_{m = 1}^M \sum\limits_{n = 1}^N {\sum\limits_{k = 1}^K {\overline{R_{m,n}^k}} }\\
&~~~~~~~~~s.t.~~c3,~c4,~c5~\text{and}~c6
\end{aligned}
\end{equation}
We define $\overline{P}=\left( P_{m,n}^{k,up},P_{m,n}^{k,down} \right)$ and the throughput expressions can be transformed to:
\begin{equation}
\overline{R_{m,n}^{k,up}}=\log \left(1+\frac{P_{m,n}^{k,up} \left| h_{m,b}^k \right|^2}{\sigma_D^2+\sigma_B^2} \right) \\
\end{equation}
\begin{equation}
\overline{R_{b,n}^{k,down}}=\log \left( 1+\frac{P_{m,n}^{k,down} \left| h_{b,n}^k \right|^2}{P_{m,n}^{k,up} \left| h_{m,n}^k \right|^2+\sigma_N^2} \right)
\end{equation}
\begin{equation}
\overline{R_{m,n}^k}=\overline{R_{m,n}^{k,up}}+\overline{R_{b,n}^{k,down}}
\end{equation}
In this case, the dual function can be written as:

\begin{equation}
g(\lambda)=\mathop{\max}\limits_{X,\overline{P}}L(\overline{P},\lambda)
\end{equation}
Where the Lagarange function and dual vector $\lambda$ is given by:
\begin{equation}
\begin{aligned}
&L(\overline{P},\lambda)=\sum\limits_{m = 1}^M \sum\limits_{n = 1}^N \sum\limits_{k = 1}^K (\overline{R_{m,n}^k}-\lambda_m P_{m,n}^{k,up}-\lambda_b P_{m,n}^{k,down})\\
&~~~~~~~~~~~~~ +\sum_{m=1}^M\lambda_m P_m
  +\lambda_bP_b\\
&\lambda=\{\lambda_1,...,\lambda_M,\lambda_b\}
\end{aligned}
\end{equation}
In this case, (P2) can be solved by solving its dual optimization problem,

\begin{equation}
\begin{aligned}
&\mathop{\min}\limits_{\lambda}~~g(\lambda)\\
&~~s.t.~~ \lambda \geq 0
\end{aligned}
\end{equation}
(19) can be solved by firstly solving (17) at each given dual point and updating the dual function using its subgradient. The subgradient of $g(\lambda)$ can be obtained by using a similar method as mentioned in \cite{11}. And the dual variables can be updated based on the expression (20)-(21), where $[\bullet]^{+}$ denotes $\max(0,\bullet)$ and the step size $\pi^{(l)}$ follows the diminishing policy in \cite{12}, i.e., $\pi^{(l)}=\pi^{(0)}/\sqrt{l}$.

\begin{equation}
\lambda_b^{(l+1)}=\left[\lambda_b^{(l)}-\pi^{(l)}\left( P_b-\sum_{m=1}^M\sum_{n=1}^N\sum_{k=1}^KP_{m,n}^{k,down} \right)\right]^{+} \\\\
\end{equation}

\begin{equation}
\lambda_m^{(l+1)}=\left[\lambda_m^{(l)}-\pi^{(l)}\left( P_m\\-\sum_{n=1}^N\sum_{k=1}^KP_{m,n}^{k,up}\right)\right]^{+} \quad\forall m \\\\
\end{equation}

To compute the dual function at each given dual point, we need to find the optimal mapping matrix $X$ and the optimal power allocation vector $\overline{P}$. In detail, an optimal power allocation scheme is derived given each possible mapping pair $(m,n,k)$, with the solution to power allocation we can do 3D mapping using the scheme proposed in Section III.

After fixing the mapping pair $(m,n,k)$ and defining equivalent channel gains $a_{m,b}^k=\frac{\left| h_{m,b}^k \right|^2}{\sigma_D^2+\sigma_B^2}$, $a_{b,n}^k=\frac{\left| h_{b,n}^k \right|^2}{\sigma_N^2}$ and $a_{m,n}^k=\frac{ \left| h_{m,n}^k \right|^2}{\sigma_N^2}$, the power allocation problem is rewritten as:

\begin{equation}
\begin{aligned}
&P3:\mathop {\max }\limits_{\left\{ \overline{P} \right\}}~~
{\log(1+P_{m,n}^{k,up}a_{m,b}^k)+\log(1+\frac{P_{m,n}^{k,down}a_{b,n}^k}{P_{m,n}^{k,up}a_{m,n}^k+1})}\\
&\qquad \qquad \ \quad -\lambda_m P_{m,n}^{k,up}-\lambda_bP_{m,n}^{k,down}\\
&\qquad~~  \ s.t.~~ \overline{P}=\left[P_{m,n}^{k,up},P_{m,n}^{k,down}\right]\geq 0
\end{aligned}
\end{equation}
With (22), by applying Karush-Kuhn-Tucker (KKT) conditions and considering different feasible regions, we can obtain the optimal power allocation scheme proved in appendix A in \cite{13}.

\section{Numerical Results}
In this section, we present the simulation results to demonstrate the performance of the proposed algorithm. We consider
a cell with radius 200 m, 8 UUE and 8 DUE are generated randomly within the same cell. To simulate practical channel
propagation, the LTE typical urban channel model is employed. The spectral density of noise is $-126$ dBm/Hz and the total bandwidth is $180$ KHz shared by $64$ subchannels. The self-interference is modeled as AWGN, with the power 3 dB larger than the noise power. And the peak power constraints for all the uplink users are the same and set to be 5 dB lower than the maximum BS transmit power.

\begin{figure}[h]
 \centering
  \includegraphics[width=9cm]{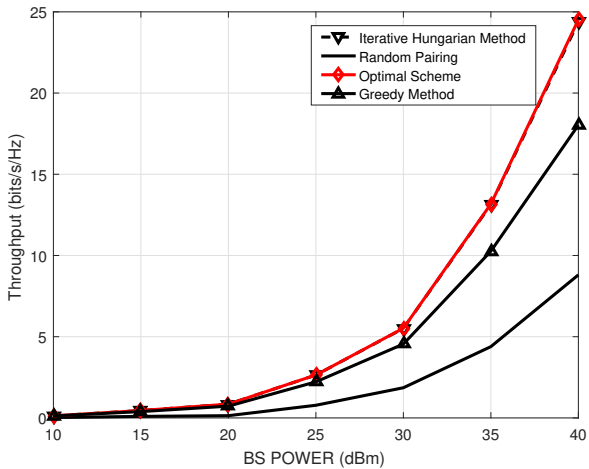}\\
  \caption{Performance comparison between different binary assignment schemes.}\label{PCF}
\end{figure}

\begin{figure}[h]
 \centering
  \includegraphics[width=9cm]{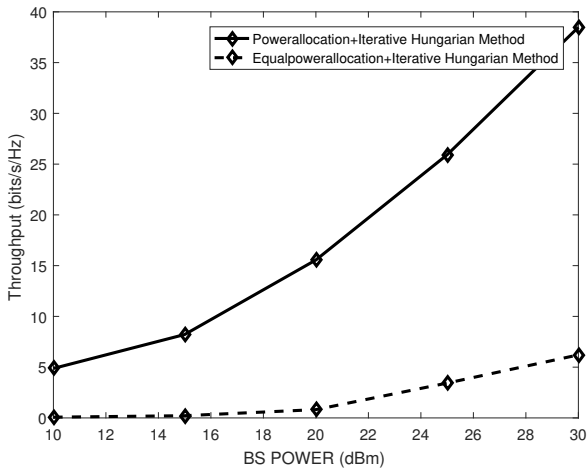}\\
  \caption{Performance of proposed joint resource allocation scheme compared with equal power allocation.}\label{PCF}
\end{figure}
%

\subsection{Binary Assignment Scheme Comparison}

With the maximum transmit power of UUE to be equal power allocation scheme, compare our proposed 3D mapping scheme with the following three benchmark schemes,

1) Exhaustive searching scheme, which is optimal.

2) Random mapping scheme, which generates a random 3D mapping in each iteration.

3) Greedy algorithm, each UUE $m$ select the DUE-subchannel pair $(n,k)$, which can maximize the throughput of pair $(m,n,k)$. Once the pair $(n,r)$ is selected, the other UUE can't select it any more.

As shown in Fig.1, with BS peak power constraint varying from 10 dBm to 30 dBm, the gaps between other approaches and the optimal scheme are quite large. In comparisons, our proposed 3D mapping scheme has the same performance with the optimal scheme. As the BS power continues raising, our proposed methodology keeps pace with the optimal case and the gap between others and our method becomes increasingly larger. Obviously, these strongly proves the correctness and the advantage of the proposed scheme.
\subsection{Joint Scheme Comparison}

In Fig.2, we compare our proposed allocation scheme with the equal power allocation one. Obviously, our scheme significantly enhance the system throughput and the performance benefits become larger as the BS power increases. This is because when we allocate more power on the high quality links, the system throughput will be better conditioned on a fixed amount of energy.


\section{Conclusion}
In this paper, we propose a joint power allocation and 3D mapping scheme to maximize a FD-BS aided OFDMA system throughput. Specially, we first decompose the joint optimization problem into two sub ones. Then resolving them by dual method and iterative Hungary algorithm sequently. Finally, sub-gradient method is applied and the optimal solution can be obtained. The numerical results demonstrate the correctness and advantages of our proposed scheme.

\balance

\end{document}